\documentclass[doublespacing]{elsart}

\usepackage{graphicx}
\usepackage{amssymb}
\usepackage{lineno}
\newcommand{\tn}{\textnormal}

\begin{document}
\begin{frontmatter}

\title{Single crystal growth and anisotropy of CeRuPO}

\author{C. Krellner} and
\ead{krellner@cpfs.mpg.de}
\author{C. Geibel}
\address{Max Planck Institute for Chemical Physics of Solids, N\"othnitzer Str. 40, D-01187 Dresden, Germany}

\begin{abstract}
We report on the single crystal growth of the ferromagnetic Kondo lattice system CeRuPO using a Sn flux method. Magnetic susceptibility and electrical resistivity measurements indicate strong anisotropy of this structurally layered compound. They evidence that the magnetic moments order ferromagnetically along the $c$-direction of the tetragonal unit cell, whereas the crystal electric field (CEF) anisotropy favors the $ab$-plane. Therefore, CeRuPO presents the unusual case within rare earth systems, where the anisotropy of the interionic exchange interaction overcomes the single ion anisotropy due to the CEF interaction.
\end{abstract}

\end{frontmatter}

\section{\label{Intro}Introduction}
Intermetallic Kondo lattice systems have attracted considerable attention in the last decades. While many Ce-based Kondo lattices show antiferromagnetic (AFM) ground states, only very few systems are known with ferromagnetic (FM) order and pronounced Kondo effects. Therefore, new FM Kondo lattice systems have to be found to study the underlying physics in more detail. Recently, we reported on the physical properties of polycrystalline CeRuPO, \cite{Krellner:2007} which seems to be one of the rare example of a FM Kondo lattice system ($T_{\rm K}\sim 10$\,K, $T_{\rm C}=15$\,K). This compound, at the border between intermetallic and oxide compounds, crystallizes with the tetragonal ZrCuSiAs type structure, which consists of alternating layers of RuP$_4$ and OCe$_4$ tetrahedra. \cite{Zimmer:1995} The results of LDA calculations indicate a strong two-dimensional (2D) character of the Fermi surface, therefore the anisotropy of the magnetic and electronic properties are of high interest. However, the synthesis of this material is difficult, due to the high vapor pressure of phosphorus and the control of oxygen.

The preparation of high quality single crystals is one of the key issues to study in detail the anisotropy and to understand the underlying physical interactions. In this contribution we will present the growth of CeRuPO single crystals from Sn flux and the study of the magnetic anisotropy by means of magnetic susceptibility $\chi(T)$ and electrical resistivity $\rho(T)$ measurements.

Two different kinds of anisotropies determines the magnetic properties of a solid. The most obvious one is that related to the different $x$, $y$ and $z$ components of the magnetic moment. It induced e.g. an anisotropy with respect to the direction of an applied external field and is therefore comparably easy to address. The second one is connected with different exchanges along different directions within the crystal structure, and can generally only be investigated with $Q$ (momentum) dependent methods, like e.g. neutron scattering. In the present paper we shall only address the former one. In magnetic rare earth systems, this anisotropy is usually determined by the single ion anisotropy induced by the crystal electric field (CEF), which is generally quite strong and lead to a pronounced anisotropy of the magnetic susceptibility. In the case of Ce$^{3+}$ in a tetragonal environment, the CEF splits the spin orbit $J=\frac{5}{2}$ multiplet into 3 Kramers doublets, which can present either a strong Ising type, an easy plane or a Heisenberg behavior depending on the CEF parameters. In contrast, the contribution to this anisotropy originating from difference in the exchange parameters for the different $x$, $y$, and $z$ component of the magnetic moment, is usually negligible in these rare earth magnetic systems. Our results demonstrate that in CeRuPO this contribution is larger than the effect of the CEF, and determines the orientation of the spontaneous FM moment.

\section{\label{ExpDet}Experimental procedure}
\begin{figure}
\includegraphics[width=8.5cm]{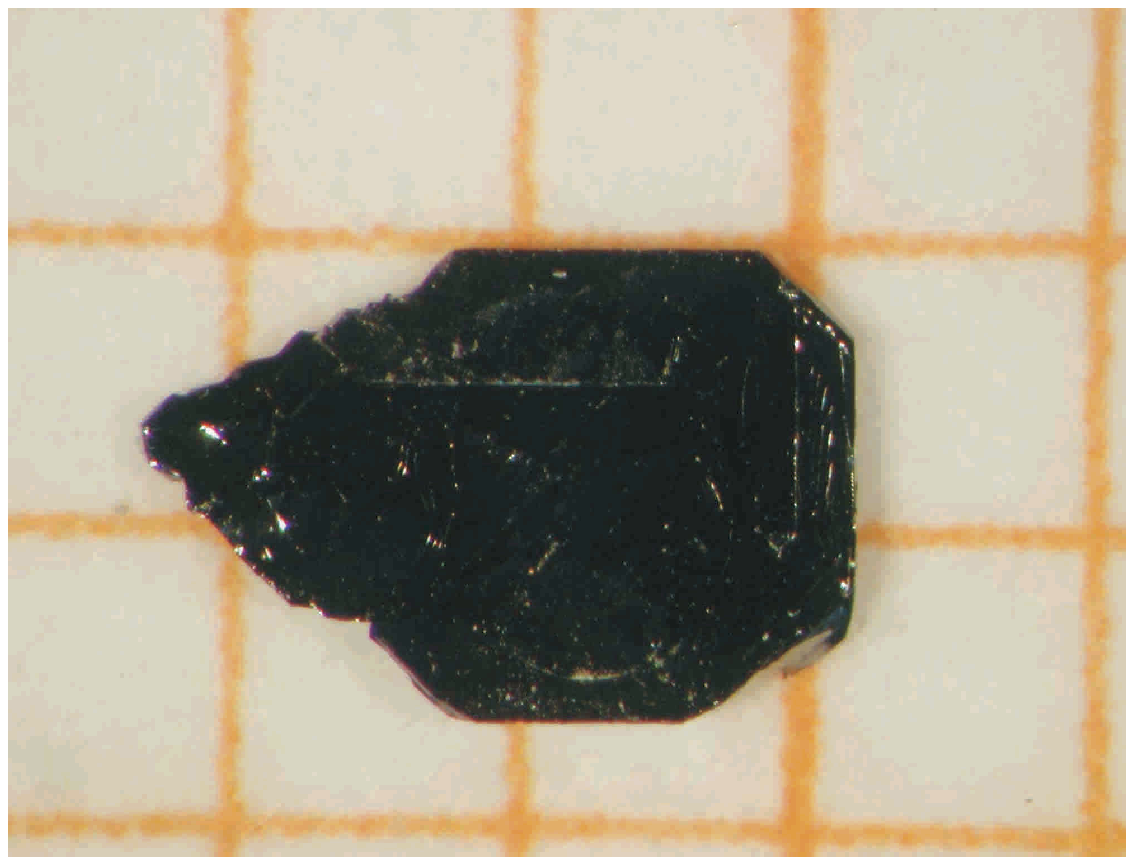}
 \caption{\label{FigPic} (Color online) Photograph of a CeRuPO single crystal (length scale 1\,mm). The direction perpendicular to the surface is parallel to the crystallographic $c$-direction.}
\end{figure}

Recently, we described the preparation of polycrystalline (PC) samples using a Sn-flux method in evacuated quartz tubes. \cite{Krellner:2007},\cite{Kanatzidis:2005} With quartz tubes the temperature range is restricted to 1000$^{\circ}$C. It turned out that within this temperature it is not possible to get single crystals (SC) large enough for physical characterization. One reason might be that the liquidus temperature of CeRuPO, at the formerly used 82\,at$\%$ Sn-concentration, is above 1000$^{\circ}$C. Therefore, we slightly changed the preparation route to achieve higher temperatures. First, we prereact the phosphorus with half of the Sn flux at 600$^{\circ}$C for 3 hours in an Al$_2$O$_3$ crucible, sealed inside an evacuated quartz tube. After quenching to room temperature, we opened the quartz tube and filled on top of the prereacted Sn-P the Ce, RuO2 and the other half of the Sn with an total molar ratio of 9:6:4.5:80.5 (Ce:P:RuO$_2$:Sn). The Al$_2$O$_3$ crucible was put inside a Ta-crucible, which was closed under an inert atmosphere. This crucible was heated with a rate of 300$^{\circ}$C per hour to 1500$^{\circ}$C and kept at this temperature for 1 hour, subsequently the melt was cooled down to 900$^{\circ}$C by moving the crucible out of the furnace with an average cooling rate of 4$^{\circ}$C per hour. After the reaction the excess Sn was dissolved in dilute HCl, leaving platelets with a surface $\sim 3$\,mm$^2$. A typical SC platelet of CeRuPO is shown in Fig.~\ref{FigPic}. The direction perpendicular to the surface corresponds to the crystallographic $c$-direction of the tetragonal unit cell, as shown by a Laue picture (see inset of Fig.~\ref{FigXray}). As a side product, we also got SC of CeRu$_2$P$_2$ and Ce$_3$Ru$_4$Sn$_{13}$, the physical properties of these new Ce-based systems will be discussed elsewhere. \cite{Krellner:2007a} 

\begin{figure}
\includegraphics[width=8.5cm]{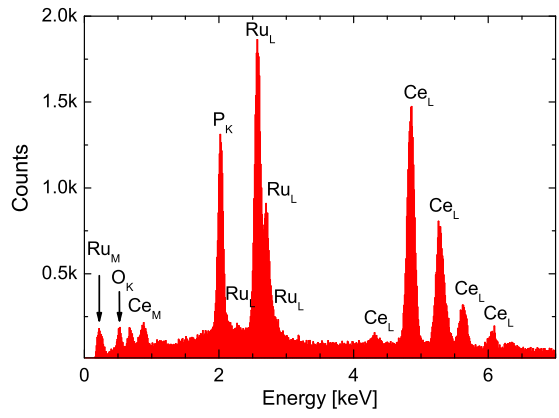}
 \caption{\label{FigEDX} (Color online) Energy dispersive X-ray spectrum of a CeRuPO single crystal.}
\end{figure}

\begin{figure}
\includegraphics[width=8.5cm]{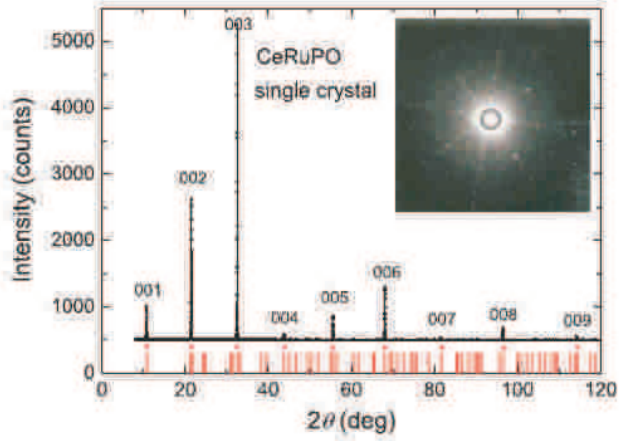}
 \caption{\label{FigXray} (Color online) X-ray diffraction pattern of a CeRuPO single crystal, in reflection mode. The bars denote the theoretical peak position \cite{Zimmer:1995}, stars indicate the $00l$ peaks. The inset shows a Laue picture of a CeRuPO single crystal, the surface is perpendicular to the $c$-axis of the tetragonal unit cell.}
\end{figure}

Several SC were investigated with electron-microprobe and energy dispersive X-ray (EDX) analysis, no foreign phases could be detected. A typical EDX spectrum is shown in Fig.~\ref{FigEDX}. All lines could be assigned to Ce, Ru, P, or O and a quantitative analysis reveals a stoichiometric Ce:Ru:P content. The presence of oxygen is confirmed by the K$_{\alpha}$ peak at 0.53\,keV; however, a quantitative analysis of this oxygen peak failed because the contribution to the spectrum is too weak, compared to the heavier elements.

X-ray diffraction (XRD) patterns were recorded on a Stoe diffractometer in both transmission and reflection mode using a monochromated Cu-K$_{\alpha}$ radiation ($\lambda = 1.5406$\,\AA). Magnetic measurements were performed in the temperature range from $2-300$\,K in a commercial Quantum Design (QD) magnetic property measurement system (MPMS) equipped with an RSO option. AC-transport resistivity measurements were carried out in a standard four-probe geometry using a commercial QD physical property measurement system (PPMS).

\section{\label{Results} Results and Discussion}
To determine the crystal lattice parameters of the SC, we performed XRD measurements in transmission mode on 4 powdered SC. The intensity of the resulting XRD pattern was low, due to the small amount of powdered SC. The lattice parameters refined by simple least square fitting, $a=4.027(3)$\,\AA $ $ and $c=8.26(1)$\,\AA $ $ are in good agreement with the reported single crystal data \cite{Zimmer:1995} and also with our PC samples. \cite{Krellner:2007} However, the accuracy of this fit was poor, indicated by the larger errors of the calculated lattice parameters. Therefore, we performed a reflection experiment on a SC to determine the $c$ parameter more accurately. The idea behind this experiment was to check the oxygen occupation between the Ce layers, because it was shown that the introduction of hydrogen in the closely related CeFeSi structure leads to a large change of the $c$ parameter. \cite{Chevalier:2005} Fig.~\ref{FigXray} shows the $00l$ reflections of a CeRuPO single crystal measured in reflection mode, the bars (red) denote the theoretical peak positions from Zimmer \textit{et al.}, \cite{Zimmer:1995} the $00l$ peaks are marked with a star. It is obvious that even at higher angles the observed $00l$ peak positions coincide with the data from Ref. \cite{Zimmer:1995}. The calculated $c$ from the higher order peaks is $c_{\rm SC}=8.259(1)$\,\AA, and is therefore only slightly larger than the values from the PC samples, where we found $c_{\rm PC}=8.256(2)$\,\AA. Compared to the large difference ($\sim 15\%$) of the $c$ parameter going from clean CeFeSi to hydrogenated CeFeSiH, we conclude no large difference in the oxygen occupancy between PC samples and SC in CeRuPO. In PC samples we have shown a stoichiometric oxygen occupancy using carrier gas-hot extraction, thus our present analysis evidences also a stoichiometric oxygen occupancy in the SC of CeRuPO.

\begin{figure}
\includegraphics[width=8.5cm]{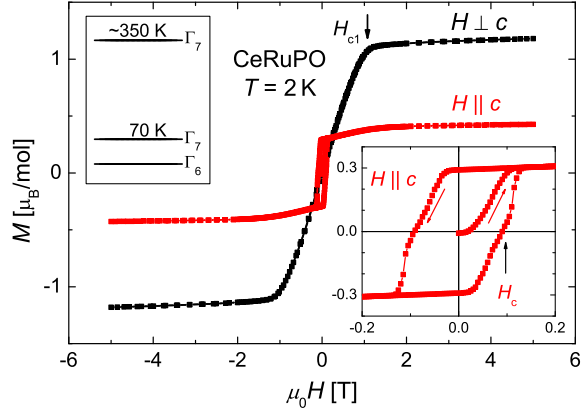}
 \caption{\label{FigMvH} Isothermal magnetization as a function of applied magnetic field at $T=2$\,K for $H\perp c$ (black symbols) and $H\parallel c$ (red symbols). The lower inset shows the well defined hysteresis curve for $H\parallel c$. The upper inset presents the proposed CEF energy level scheme.}
\end{figure}

To study the magnetic anisotropy of CeRuPO, we have performed dc-magneti-zation measurements on a SC for $H\perp c$ and $H\parallel c$. 
A large anisotropy is evident from Fig.~\ref{FigMvH}, where we show the magnetization measurements in the ordered phase at 2\,K for both field directions. For $H\perp c$ the magnetization increases linearly with a slope of $1.1\,\mu_{\rm B}/$T at small fields and saturates above $\mu_0H_{c_1}=1$\,T with an absolute value of $\mu_{\rm sat}^{ab}=1.2\,\mu_{\rm B}$. For $H\parallel c$ we observe a spontaneous magnetization at small fields, followed by a small linear increase (which might be due to small misalignment), saturating also above $H_{c_1}$ with a magnetic moment of $\mu_{\rm sat}^c=0.43\,\mu_{\rm B}$. In the inset of Fig.~\ref{FigMvH} we have plotted the well defined hysteresis loop, together with the virgin curve for $H\parallel c$ with a spontaneous magnetic moment of  $0.3\,\mu_{\rm B}$ and a coercive field of $H_c=0.1$\,T. For $H\perp c$ we could not observe any hysteresis. This behavior strongly suggests that CeRuPO is a collinear FM with the moments aligned along the $c$-direction, but with the easy magnetization direction perpendicular to the $c$-axis. The observed behavior is most likely due to a competition between the CEF anisotropy and the anisotropy of the Ruderman-Kittel-Kasuya-Yosida (RKKY) exchange interaction with respect to the $x$, $y$ and $z$ component of the magnetic moment. A similar behavior was also observed and studied in detail by means of neutron and magnetization measurements on the FM Kondo system YbNiSn by Bonville \textit{et al.} \cite{Bonville:1992} On the one hand the magnetic exchange interaction (RKKY), which is responsible for the FM order, is larger for the $z$ component of the moment than for the $x$ and $y$ component, yielding a spontaneous FM ordering of the moments along the $c$-direction. On the other hand, the CEF anisotropy leads to a ground state CEF doublet with a larger saturation moment in the basal plane than along the $c$-axis. This scenario explains both the absence of hysteresis as well as the linear increase of the magnetization for $H\perp c$, which is due to the rotation of the FM moments towards the $ab$-plane. Above $H_{c1}$ the FM moments are completely aligned along the $ab$-plane with the full moment of the CEF ground state doublet for this direction. An alternative scenario invoking a canted AFM ordering with AFM moments slightly canted away from the $ab$-plane can be excluded, because of the presence of inversion symmetry in the structure, which prohibits a macroscopic net FM canted moment. 

Now we discuss the magnetization data on the basis of the CEF model, whose Hamiltonian for the $J=5/2$ multiplet of the Ce$^{3+}$ ion located in tetragonal point symmetry can be written as
\begin{equation}
H_{\rm CEF}=B_2^0O_2^0+B_4^0O_4^0+B_4^4O_4^4\tn{ ,}
\label{EqHCEF}
\end{equation}
where $B_n^m$ are the Stevens coefficients and $O_n^m$ are the Stevens equivalent operators. \cite{Stevens:1952} The resulting three doublet eigenstates can be expressed with a single mixing coefficient $\eta$ \cite{Ohama:1995}
\begin{eqnarray}
\Gamma_6 & \quad:	&  	\quad\bigl{|} \pm \frac{1}{2}\bigr{\rangle} \\
\Gamma_7&^{(1)}  :	&  	\quad\eta\,\bigl{|}\pm\frac{5}{2}\bigr{\rangle} + \sqrt{1-\eta^2}\,\bigl{|}\mp\frac{3}{2}\bigr{\rangle} \\
\Gamma_7&^{(2)}  :	&  	\quad\sqrt{1-\eta^2}\,\bigl{|}\pm\frac{5}{2}\bigr{\rangle} - \eta\,\bigl{|}\mp\frac{3}{2}\bigr{\rangle}\tn{ .} 
\label{Gamma}
\end{eqnarray}
It is now possible to determine the ground state wave function by comparing the theoretical values of the saturated moment with the experimental data for the two different directions. We found that the ground state is the $\Gamma_6$ wave function and the excited states are the two $\Gamma_7$ eigenstates, because for this configuration the saturation moment along $z$ is $\mu_{\rm sat}^z=\frac{1}{2}g_{\rm L}\mu_{\rm B}\sim0.43\,\mu_{\rm B}$ and $\mu_{\rm sat}^x=\frac{3}{2}g_{\rm L}\mu_{\rm B}\sim1.29\,\mu_{\rm B}$ along $x$, ($g_L=\frac{6}{7}$ is the Land\'e factor for Ce$^{3+}$) which is in very good agreement with the experimental obtained values for the two different field directions. If one assumes one of the $\Gamma_7$ as the ground state, the maximal value of $\mu_{\rm sat}^x=\frac{1}{2}\sqrt{5}g_{\rm L}\mu_{\rm B}\sim0.96\,\mu_{\rm B}$ for $\eta=1/\sqrt{2}$ is well below the experimental $\mu_{\rm sat}^{ab}$ and rule out such a configuration. From the temperature dependence of the entropy we have shown that the two excited Kramer's doublets are 6 and 30\,meV above the ground state. \cite{Krellner:2007} Therefore, we can draw a preliminary CEF level scheme, shown in the inset of Fig.~\ref{FigMvH}. However, we can not determine the exact values of the Stevens coefficients from Eq.~\ref{EqHCEF}, which would need a more sophisticated analysis and/or further experiments e.g. inelastic neutron measurements. 

\begin{figure}
\includegraphics[width=8.5cm]{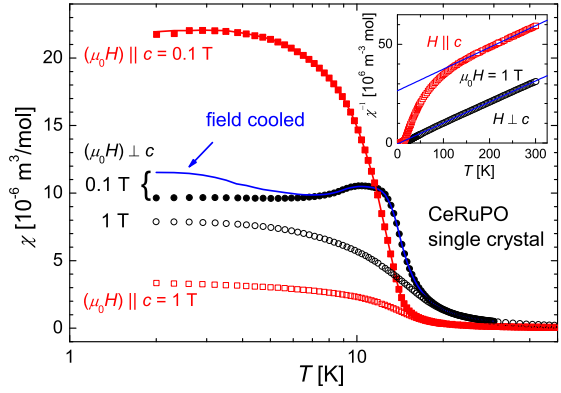}
 \caption{\label{FigMvT} Magnetic susceptibility vs. temperature at $\mu_0H=0.1$\,T (closed symbols) and 1\,T  (open symbols) for $H\parallel c$ (red) and $H\perp c$ (black). The solid line (blue) presents a field cooled experiment for $H\perp c$ at 0.1\,T. $\chi^{-1}(T)$ for both directions is shown in the inset, together with the Curie Weiss fits for $T>150$\,K.}
\end{figure}

The complex anisotropic behavior of CeRuPO is also reflected in the temperature dependence of the susceptibility $\chi(T)$ measured along both crystallographic directions shown in Fig.~\ref{FigMvT}. $\chi(T)$ for $H\parallel c$ presents the typical behavior of a FM system. The ordering below $T_{\rm C}=14$\,K is clearly visible in a strong increase of the susceptibility, reaching a constant value below 5\,K. This increase is much weaker for $\mu_0H=1$\,T. For $H\perp c$ the magnetic field dependence of $\chi(T)$ is not so pronounced. At $\mu_0H=0.1$\,T a small peak appears below $T_{\rm C}$, similar to what we have observed in PC samples. \cite{Krellner:2007} The precise physical origin of this antiferromagnetic signature is still unclear, but we think that it is due to the complex anisotropy, because a similar feature was also observed in YbNiSn for a direction perpendicular to the FM axis (see Fig. 7 in Kasaya \textit{et al.} \cite{Kasaya:1991}). We also have performed field cooled experiments for both directions of the magnetic field at $\mu_0H=0.1$\,T (solid lines in Fig.~\ref{FigMvT}), but only for $\mu_0H\perp c$ we could observe a small difference at low field, most probably due to domain reorientation. At high temperature ($T>150$\,K) $\chi^{-1}(T)$ follows a Curie-Weiss law with $\Theta^c_{\rm W} \simeq -250$\,K, $\Theta^{ab}_{\rm W} \simeq +4.2$\,K, and effective magnetic moments $\mu_{\rm eff}=2.4\,\mu_{\rm B}$ for both directions, close to that of the free ion moment ($2.54\,\mu_{\rm B}/\rm{Ce}^{3+}$, see inset of Fig.~\ref{FigMvT}). The anisotropy of $\Theta_{\rm W}$ confirms the conclusions drawn above, that the CEF anisotropy favors the spins to be perpendicular to the $c$-axis. The large difference between $\Theta^{ab}_{\rm W}$ and $\Theta^c_{\rm W}$ implies a large positive $B_2^0$ CEF coefficient, which is expected to result in a $\Gamma_6$ CEF ground state, in agreement with our deduction based on the magnetization data at 2\,K. At lower temperature $\chi^{-1}(T)$ changes slope and the data can be fitted with a positive Weiss temperature of the order of $T_{\rm C}$ for both directions.

\begin{figure}
\includegraphics[width=8.5cm]{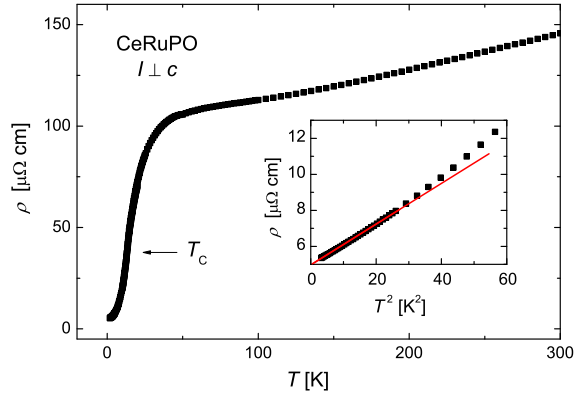}
 \caption{\label{FigRho} (Color online) Temperature dependence of the electrical resistivity of CeRuPO with the current perpendicular to the $c$ direction. The low temperature part is shown in the inset as $\rho$ vs. $T^2$, below 5\,K $\rho\propto T^2$ indicates the formation of a Fermi-liquid ground state.}
\end{figure}

The temperature dependence of the resistivity for the current flowing perpendicular to the $c$-axis is plotted in Fig.~\ref{FigRho}. Measurements with current along the $c$-axis could not yet be performed because of the small thickness of the platelets. The overall temperature dependence is very similar to what we have reported on PC samples. \cite{Krellner:2007} Three different temperature regions can be identified: (I) Above 50\,K the sample shows a linear behavior typical of a conventional metal. (II) Below 50\,K there is a pronounced decrease, which is a distinct feature of a Kondo lattice system and can be explained by the onset of coherent Kondo scattering, due to the hybridization between the localized $4f$ and the conduction electrons. (III) In contrast to the PC samples, in the ordered state $\rho(T)$ follows a power law dependence with $\rho\propto T^2$, indicating the presence of a Landau Fermi liquid. In the inset of Fig.~\ref{FigRho} we have plotted this dependence as $\rho$ vs. $T^2$ and the linear fit (red curve) gives $\rho(T^2)=\rho_0+AT^2$, with $\rho_0=5\,\mu\Omega$cm and $A=0.11\,\mu\Omega$cmK$^{-2}$.

\begin{figure}
\includegraphics[width=8.5cm]{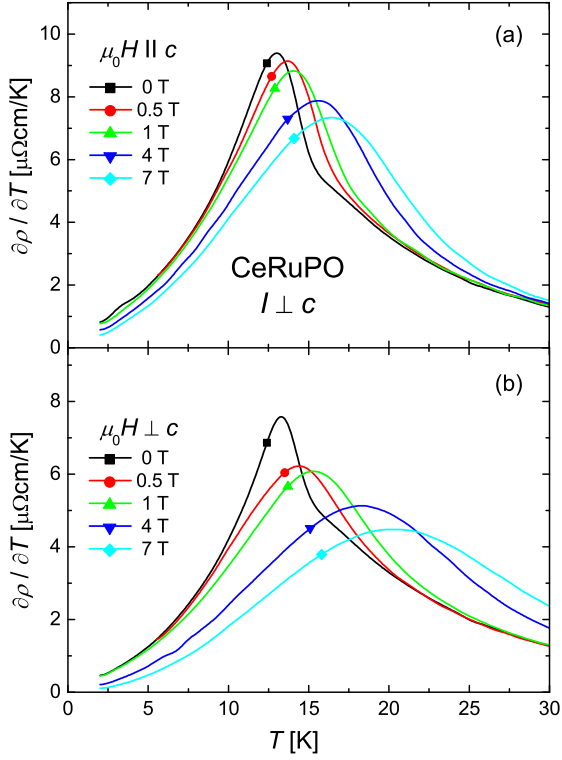}
 \caption{\label{FigdRho} (Color online) Derivative of the resistivity of CeRuPO with respect to temperature plotted as $\partial \rho / \partial T$ vs. $T$. for two different samples with $H\parallel c$ (a) and $H\perp c$ (b). The anomaly due to the onset of FM order is clearly visible as a sharp peak. It shifts to higher temperatures with applied magnetic field.}
\end{figure}

The anomaly in the resistivity at the FM ordering temperature is visible in Fig.~\ref{FigdRho}, where the temperature dependence of the derivative of the resistivity, $\partial\rho/\partial T$ vs. $T$ is plotted for two different samples with the magnetic field parallel and perpendicular to the $c$-direction. We have shown previously on the PC samples that $\partial\rho/\partial T(T,H)$ behaves similar to the temperature and field dependence of the specific heat and can be therefore taken as a direct measure of the development of the FM order. In the upper panel of Fig.~\ref{FigdRho} the magnetic field is applied along the $c$-direction, whereas current is flowing within the $ab$-plane. For $\mu_0H=0$, $\partial\rho/\partial T$ present a sharp peak at $T=13$\,K which broadens and shifts to higher temperatures with increasing magnetic field. For $\mu_0H=7$\,T the peak is situated at 16.5\,K. The field dependence is stronger for field applied within the $ab$-plane (see lower panel of Fig.~\ref{FigdRho}). There the peak shifts from 13\,K at zero field to 20\,K for $\mu_0H=7$\,T. An increase of $T_{\rm C}$ with magnetic field is expected for a FM, since the external field is supporting the exchange field. The stronger increase for field along the basal plane compared to field along the $c$-axis can be related to the higher saturation moment in the basal plane, which leads to a larger free energy gain in the external magnetic field. 

We now discuss the small difference in $T_{\rm C}$ of the SC compared to our previous study on PC samples ($T_{\rm C}^{\rm PC}=15$\,K). \cite{Krellner:2007} The $T_{\rm C}$ of the SC ($T_{\rm C}^{\rm SC}=14$\,K) was determined from a Curie Weiss Fit of the $\chi^{-1}(T)$ data below 30\,K. The anomaly in $\partial\rho/\partial T (T)$ confirms this $T_{\rm C}^{\rm SC}$, if one takes into account the asymmetric shape of the peak (see Fig.~\ref{FigdRho}). However, the peak is broader in the SC data compared to the PC measurement, and therefore it is more difficult to determine $T_{\rm C}^{\rm SC}$. The specific heat data for a SC gives similar results (not shown), the peak at $T_{\rm C}^{\rm SC}=14$\,K is rather broad. The reduced ordering temperature and the broader anomalies at $T_{\rm C}^{\rm SC}$ indicate that the crystallinity of the PC material is better than in the here presented SC. This is also evident from a smaller residual resistivity ratio for the SC $\rho_{300\,\rm K}^{\rm SC}/\rho_0^{\rm SC}=30$ compared to $\rho_{300\,\rm K}^{\rm PC}/\rho_0^{\rm PC}=50$ in the PC sample. However, we can exclude a large difference in the oxygen occupancy for SC and PC, as we have shown above. Thus the broader transition at a lower $T_{\rm C}$ is likely due to a larger defect density, like e.g. a P-O substitution.

\section{\label{Concl}Conclusions}
In summary, we succeeded in growing single crystals large enough to determine the anisotropy of CeRuPO by means of magnetic susceptibility and resistivity measurements. We showed that the spontaneous FM moments point towards the $c$-axis, while the saturation moment is larger in the basal plane. The latter property indicates an easy plane single ion CEF ground state, while the former property implies a stronger exchange for the $z$ component of the moment than for the $x$ and $y$ component, this anisotropy overcoming that of the CEF (at low $T$). A more detailed analysis gives for the CEF ground state the $\Gamma_6$ eigenstate, composed of $|\pm\frac{1}{2}\rangle$ wave functions. In addition, we have presented detailed EDX and XRD results which show that the difference in the oxygen occupancy is small for PC samples compared to the SC. However, the SC have a slightly smaller ordering temperature ($T_{\rm C}^{\rm SC}=14$\,K) compared to the PC samples ($T_{\rm C}^{\rm PC}=15$\,K). Further studies are necessary to determine the precise origin of this difference.

\section*{Acknowledgements}
The authors thank U. Burkhardt and P. Scheppan for chemical analysis of the samples, and N. Caroca-Canales and R. Weise for technical assistance in sample preparation. The Deutsche Forschungsgemeinschaft (SFB 463) is acknowledged for financial support.


\newpage

$ $
\newpage

$ $
\newpage

$ $
\newpage

$ $
\newpage

$ $
\newpage

$ $
\newpage

\end{document}